\documentclass[5p]{elsarticle}

 \usepackage{graphicx}
 \usepackage[font=footnotesize]{subfig} 

\usepackage{amssymb}
\usepackage{color}

\def\beq{\begin{equation}}
\def\eeq{\end{equation}}
\def\bea{\begin{eqnarray}}
\def\eea{\end{eqnarray}}

\newcommand*{\eqref}[1]{Eq.~(\ref{eq:#1})}
\newcommand*{\eqlab}[1]{\label{eq:#1}}
\newcommand*{\figref}[1]{Fig.~\ref{fig:#1}}
\newcommand*{\figlab}[1]{\label{fig:#1}}

\newcommand*{\seclab}[1]{\label{sec:#1}}

\def\NIM#1#2#3{Nucl.~Inst.~Meth.~\VYP{A#1}{#2}{#3}}
\def\ApP#1#2#3{Astropart.~Phys.~\VYP{#1}{#2}{#3}}
\def\VYP#1#2#3{{\bf #1}, #3 (#2)}  

\newcommand{\etal}{\mbox{\textit et al.}}                       %

\begin{document}

\begin{frontmatter}



\title{The air shower maximum probed by Cherenkov effects from radio emission}


\author[KVI]{Krijn D. de Vries}
\ead{krijndevries@gmail.com}
\author[KVI]{Olaf Scholten}
\author[SUBA]{and Klaus Werner}
\address[KVI]{Kernfysisch Versneller Instituut, University
of Groningen, 9747 AA, Groningen, The Netherlands}
\address[SUBA]{SUBATECH,
University of Nantes -- IN2P3/CNRS-- EMN,  Nantes, France}




\begin{abstract}
Radio detection of cosmic-ray-induced air showers has come to a flight the last decade. Along with the experimental efforts, several theoretical models were developed. The main radio-emission mechanisms are established to be the geomagnetic emission due to deflection of electrons and positrons in Earth's magnetic field and the charge-excess emission due to a net electron excess in the air shower front. It was only recently shown that Cherenkov effects play an important role in the radio emission from air showers. In this article we show the importance of these effects to extract quantitatively the position of the shower maximum from the radio signal, which is a sensitive measure for the mass of the initial cosmic ray. We also show that the relative magnitude of the charge-excess and geomagnetic emission changes considerably at small observer distances where Cherenkov effects apply.
\end{abstract}

\begin{keyword}
Radio detection \sep Air showers \sep Cosmic rays \sep
\sep Coherent radio emission \sep Geomagnetic Cherenkov radiation \sep Mass determination


\PACS 95.30.Gv \sep 95.55.Vj \sep 95.85.Ry \sep 96.50.S- \sep
\end{keyword}
\end{frontmatter}

\section{Introduction}
Recent experiments (LOPES~\cite{Fal05,Ape06}, CODALEMA~\cite{Ard06,Ard09}) showed that radio emission can be used as a new complementary technique for measuring cosmic-ray-induced air showers. This gave a new impulse to the field of radio emission from air showers and triggered several new experiments at the Pierre Auger Observatory~\cite{Ber07,Cop09,Rev09,AERA}, and LOFAR~\cite{LOFAR}. From the theoretical point of view several models have been developed to give a description of the radio emission from air showers, such as REAS3~\cite{REAS3}, MGMR~\cite{Sch08}, and more recently ZHAires~\cite{ZHAires}, SELFAS~\cite{SELFAS}, CoReas~\cite{Hue12}, and EVA~\cite{Wer12}. 

There used to be a huge discrepancy between the different macroscopic and microscopic models. The microscopic models in time domain like REAS~\cite{Hue05} were based on the geosynchrotron emmission of electrons and positrons gyrating in Earth's magnetic field. The macroscopic models in time domain like MGMR~\cite{dVries10} were based on the macroscopic geomagnetic current due to deflection of charged particles in Earth's magnetic field and the net negative electron excess in the shower front. Where the microscopic models in time domain used to predict uni-polar pulses, the macroscopic models predicted bi-polar pulses with a difference in magnitude up to a factor ten. This has only recently been resolved with the addition of a missing radiation component to the microscopic models~\cite{Hue10-Arena}.  

It follows that similar results are obtained for macroscopic models like MGMR~\cite{dVries10} and models using a microscopic description by adding the electric field contributions of the separate electrons like REAS3~\cite{REAS3}. From these results it can be concluded that the emission mechanisms are understood. The main geomangetic mechanism is due to the drift of charged particles in Earth's magnetic field. This gives rise to a net macroscopic current in the direction of the Lorentz force acting on the particles. The time variation of the induced vector potential gives rise to a net electromagnetic pulse. A second emission mechanism is due to the net negative charge in the shower front. The polarization of the signal as a function of the observer position with respect to the impact point of the shower can be used to disentangle the geomagnetic emission and the charge-excess emission~\cite{dVries10}. Furthermore, the Lateral Distribution Function (LDF), given by the radio signal strength as a function of antenna distance to the shower axis, can be used to give information about the composition of the cosmic-ray spectrum at the highest energies~\cite{dVries10}. 

In~\cite{Wer08} it was shown that even though the deviation of the index of refraction from unity, the refractivity, is small ($O(10^{-3})$), this deviation gives rise to Cherenkov effects at realistic observer distances. The importance of these effects were discussed in~\cite{ZHAires,Hue12,dVries11-PRL,Kal06} for a simplified shower geometry, where the LDF was shown to have a distinct peak. In~\cite{Wer12} we showed the first results of the EVA-code based on realistic charge and current distributions in the air shower which are obtained from Monte-Carlo simulations. 

In this article we show that due to Cherenkov effects it is possible to accurately extract from the radio signal the position where the shower profile reaches its maximum. This is closely related to the mass of the initial cosmic ray. In addition Cherenkov effects have a large effect on the relative magnitude of the charge-excess emission with respect to the geomagnetic emission. The calculations presented are done using the EVA~\cite{Wer12} simulations to which we give a short introduction. 

\section{The model\seclab{model}}
The basis of the EVA model lies at the Li\'enard-Wiechert potentials from classical electrodynamics~\cite{Jac-CE}. The particle distributions in the shower front are introduced by the weight function $w(\vec{r},h)$. The total vector potential is given by the convolution of the point-like potential and the weight function,
\bea
A^{\mu}_w(t,\vec{x})&=&\int d^{2}r\int dh\;w(\vec{r},h)A^{\mu}_{PL}(t,\vec{x}-\vec{\xi}),
\eqlab{vec-pot}
\eea
where the point-like vector potential given by,
\beq
A^{\beta}_{PL}(t,\vec{x}-\vec{\xi}(t'))=
\frac{\mu _0}{4\pi}\frac{J^\beta_{PL}}{|\tilde{R}V|}\Bigr{|}_{t=t'},
\eqlab{vec-pl}
\eeq
is based on the four-current
\beq
J_{PL}^{\beta}(t',\vec{x})=J^{\beta}(t')\delta^{3}(\vec{x}-\vec{\xi}(t')).
\eeq
The vector potential is to be evaluated at the negative retarded shower time $t'$. We define $\vec{\xi}(t')=-ct' \vec{e}_{x^{||}}$, where $\vec{e}_{x^{||}}$ is the unit vector pointing along the shower axis. The velocity is defined as $V=c^{-1}d\xi/dt'$. The four-vector $\tilde{R}$ is defined by the light-cone condition $\tilde{R}^0 \equiv L \equiv c(t-t')$ and $\tilde{R}^i=-L\partial/\partial\xi^i L$. Where $L$ is the optical path length between the source $\vec{\xi}(t')$ and the observer position $\vec{x}$. In~\cite{Wer08} it was already shown that the denominator in the vector potential, $\tilde{R}V$, can be linked to the derivative of the retarded time with respect to the observer time $\tilde{R}V=\tilde{R}^0/(dt'/dt)$. Furthermore, it was shown that for a realistic index of refraction the denominator could hit a zero leading to a divergence in the vector potential. This divergence is due to the well known Cherenkov effect. To overcome these divergences a partial integration is done such that the derivatives acting on the denominator $|\tilde{R}V|$ are shifted to the weight function $w(r,h)$. The final expression for the electric field is now given by~\cite{Wer12},
\bea
E^{||}&=&-c\int d^2\eta^\perp\int \limits^{h_k}_0 d\lambda\nonumber\\
&&\left\{ w' A^0_{PL}-\beta w'A^{||}+w\dot{A}^{||}_{PL}\right\}\nonumber\\
E^{\perp i}&=&c\int d^2\eta^\perp\int \limits^{h_k}_0 d\lambda\nonumber\\
&& \left\{ w^iA^0_{PL}+\beta w' A^{\perp i}_{PL}-\beta w \dot{A}^{\perp i}_{PL}\right\},
\eqlab{field-tot}
\eea
where we defined the coordinates $\vec{\eta}^\perp=\vec{x}^{\perp}+\vec{r}$ and $\lambda=h_k-h$. Here $h_k$ is defined as the critical distance behind the shower front where Cherenkov effects occur and the vector potential diverges. Looking at~\eqref{field-tot} we see that by making use of the finite extent of the particle distributions in the shower a non-singular expression is obtained. The derivatives of these distributions, $w'=dw/dh$, and $w^{i}=dw/dx^{i}$, are finite where the vector potential, $A^{\mu}_{PL}$, has a square-root divergence which is now safely integrated.

\section{The emission mechanisms}
\seclab{emission}
The main emission mechanism is due to the deflection of electrons and positrons in Earth's magnetic field. The induced drift gives rise to a net macroscopic current in the direction of the Lorentz force. The electric field is given by the time derivative of the induced vector potential and is polarized in the direction of the Lorentz force. There is an additional emission mechanism due to a net electron excess in the air shower front. This electron excess induces a scalar potential, and the electric field is given by the spatial derivatives acting on the potential. The polarization is given by the spatial derivative and hence becomes radial in the direction of the observer with respect to the shower axis~\cite{dVries09}.
Evidence for the charge-excess component in air showers has been observed at the Pierre Auger Observatory~\cite{Schoorl11} and at the CODALEMA site~\cite{Marin11}. Apart from an additional test of the Monte Carlo shower simulation, the importance of a thorough understanding of the charge-excess emission at distances where Cherenkov effects play a role lies in the fact that this is the main emission mechanism in dense media where Cherenkov effects are dominant. This emission is thus crucial for experiments searching for GZK neutrino's through radio emission from showers induced in ice~\cite{ANITA} or moon rock~\cite{Numoon}. 

The maximum for the geomagnetic radio emission occurs when the shower maximum is observed at the Cherenkov angle. The position of the shower maximum $X_{max}(\mathrm{g/cm^2})$, is defined by the depth at which the total electromagnetic component reaches its maximum. For charge-excess the maximum in the radio signal is seen when Cherenkov emission is observed from the maximum of the electron excess profile. The electron excess profile does not necessarily peak at the same height as the total shower profile and the Cherenkov peak might be observed at a different observer distance. Therefore, polarization analysis gives additional information about the relative peak position of the total electromagnetic component with respect to the charge-excess profile.

For the charge-excess analysis in~\cite{Schoorl11,Fraenkel12}, the variable $R$ defined as
\beq
R=\frac{\vec{E}_{x}\cdot\vec{E}_{y}}{(E_x^2+E_y^2)^{1/2}}\;,
\eqlab{R}
\eeq
is used. By definition the $\hat{x}$ direction is given by the projection of the geomagnetic component $\vec{e}_{geo}=-\vec{e}_{\beta}\times\vec{e}_{B}$ on the ground plane, where $\vec{e}_{\beta}$ is the unit vector pointing along the shower axis and $\vec{e}_{B}$ is the unit vector pointing in the direction of Earth's magnetic field. The $\hat{y}$ direction is given perpendicular to $\hat{x}$ in the ground plane. With this definition $R$ vanishes independent of the observer position if there would be no electric field component beside the geomagnetic emission. Since the charge-excess polarization is pointing radially inward from the observer position to the shower axis, $R$ vanishes if the observer is positioned on the positive $\hat{x}$ axis, becomes negative moving to the $\hat{y}$ axis, goes to zero again on the $-\hat{x}$ axis, and becomes positive on the $-\hat{y}$ axis.

In the following, we consider the simplified geometry of a perpendicular incoming air shower with respect to Earth's surface. The magnetic field is pointing to the North. The observer angle is defined from East ($\psi=0$~degrees) to North ($\psi=90$~degrees). For the considered geometry a sinusoidal pattern should become visible if $R$ is plotted as a function of the observer angle $\psi$. 
\begin{figure}[ht]
\centerline{
\includegraphics[width=.5\textwidth, keepaspectratio]{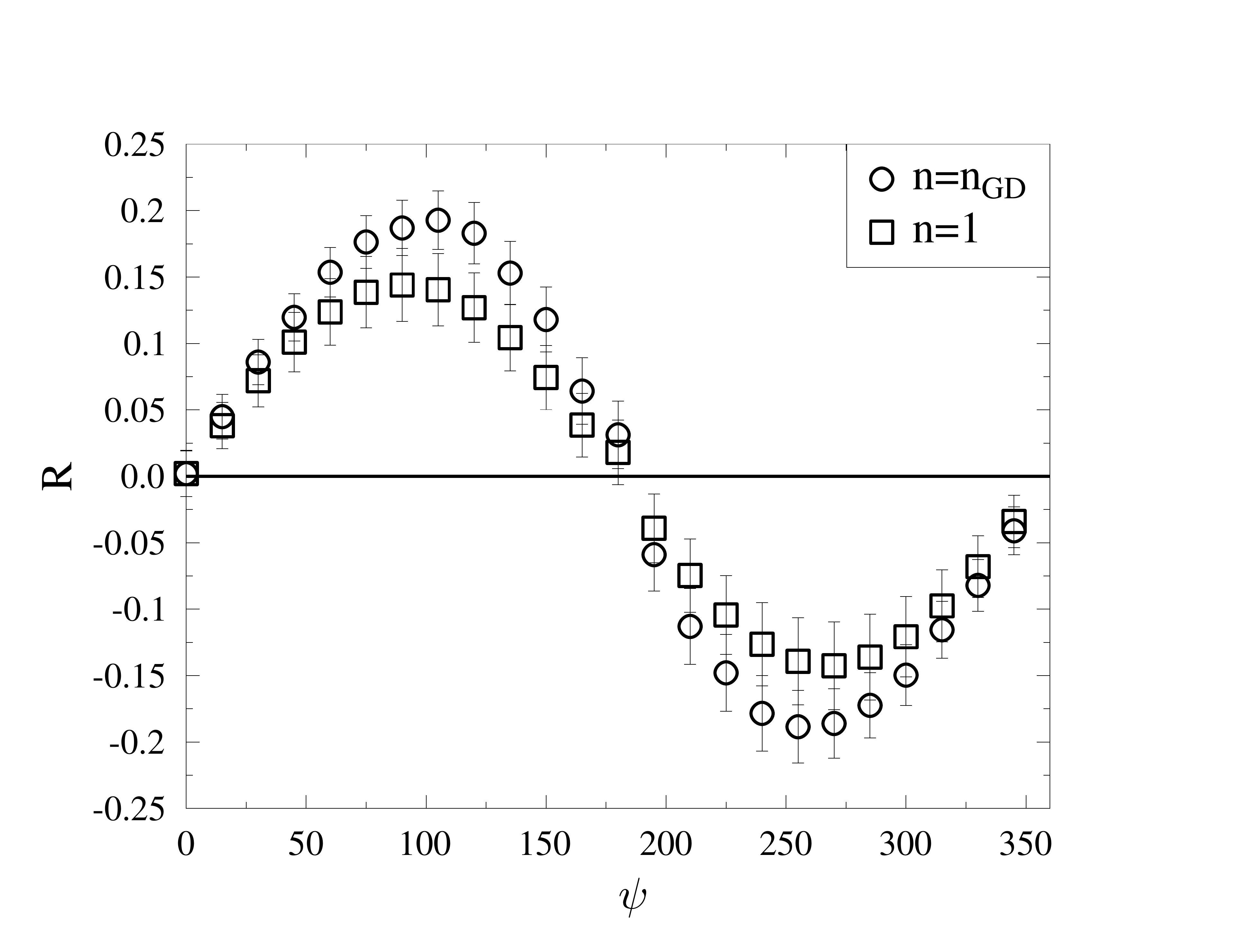}}
\caption{The average value of $R$ as defined in~\eqref{R} as a function of $\psi$ the observer angle in degrees for a set of 24, $10^{17}$~eV, showers with a proton as primary shower-inducing particle. The primary inducing shower particle moves perpendicular to Earth's surface. The squares give $R$ for an index of refraction equal to unity, the circles give $R$ for a realistic index of refraction following the law of Gladstone and Dale given in~\eqref{gd}.}
\figlab{Rvspsi}
\end{figure}
In~\figref{Rvspsi} the average value of $R$ for a set of 24, $10^{17}$~eV, showers with a proton as primary shower-inducing particle is shown. This is done for an index of refraction equal to unity as well as a more realistic index of refraction following the law of Gladstone and Dale, $n_{GD}$, given by
\begin{equation}
n_{GD}=1+0.226\mathrm{\frac{g}{cm^2}}\rho(h),
\eqlab{gd}
\end{equation}
where $\rho(h)$ is the air density at an atmospheric height $h$. The observer distance $d=75$~m is chosen in such a way that the shower maximum is observed at the Cherenkov angle for geomagnetic emission. For an index of refraction of unity we clearly observe a sinusoidal-like pattern. Since Cherenkov effects do not affect the polarization of the signal we do not expect the qualitative features of this pattern to change significantly for a realistic index of refraction. This is indeed the case as can be observed from~\figref{Rvspsi}. Nevertheless, even though the pattern does not change significantly, the maximum value of $R$ increases. 

This is also seen in~\figref{Rvsd} where we plot the absolute value of $R$ at an observer angle of 90 degrees as a function of distance from the shower axis, $|R|(90^\circ;d)$. The charge-excess emission is radially polarized and should thus vanish at the shower axis. This is not the case for the geomagnetic emission which is polarized in the direction of the Lorentz force. It follows that charge-excess emission, and thus $R$, scales with the opening angle with respect to the point of maximum emission. For an index of refraction equal to unity this is the opening angle with the shower maximum, which steadily increases with observer distance, and hence $|R|$ also steadily increases with observer distance. For a realistic index of refraction, however, the opening angle of maximum emission is equal to the Cherenkov angle and stays constant over the range of distances where Cherenkov effects are dominant. This is observed in~\figref{Rvsd} in the region below $140$ meters. Below 50 meters, $|R|$ becomes considerably larger when Cherenkov effects are included into the simulation. At slightly larger observer distances, but still in the regime where Cherenkov effects are important, $|R|$ becomes smaller when Cherenkov effects are applied. In this region something more subtle is going on.

This can be understood from~\figref{profile}, where the shower profile (full line) for the total electromagnetic component is given as well as the charge-excess profile (dashed line). It follows that the charge-excess profile peaks closer to Earth's surface ($\sim 740\;\mathrm{g/cm^2}$) than the shower profile ($\sim680\;\mathrm{g/cm^2}$). Due to this effect, at $140$ meters Cherenkov effects have diminished for the charge-excess component, but are still large for the geomagnetic emission and hence $R$ becomes small. 

If the charge-excess profile would have peaked at a larger height compared to the shower profile for the total electromagnetic component, $|R|$ as plotted in~\figref{Rvsd} would always be larger when Cherenkov effects are included into the simulation compared to the calculations done for an index of refraction equal to unity. This is a large effect, which should be measurable. The value of $|R|$ in the regime where Cherenkov effects apply is thus sensitive to the relative position of the charge-excess maximum with respect to the maximum for the shower profile given by the total electromagnetic component. 

For distances larger than $140$ meters, Cherenkov effects become small and the value of $R$ for a realistic index of refraction converges to the value of $R$ for a constant index of refraction equal to unity.
\begin{figure}[ht]
\centerline{
\includegraphics[width=.5\textwidth, keepaspectratio]{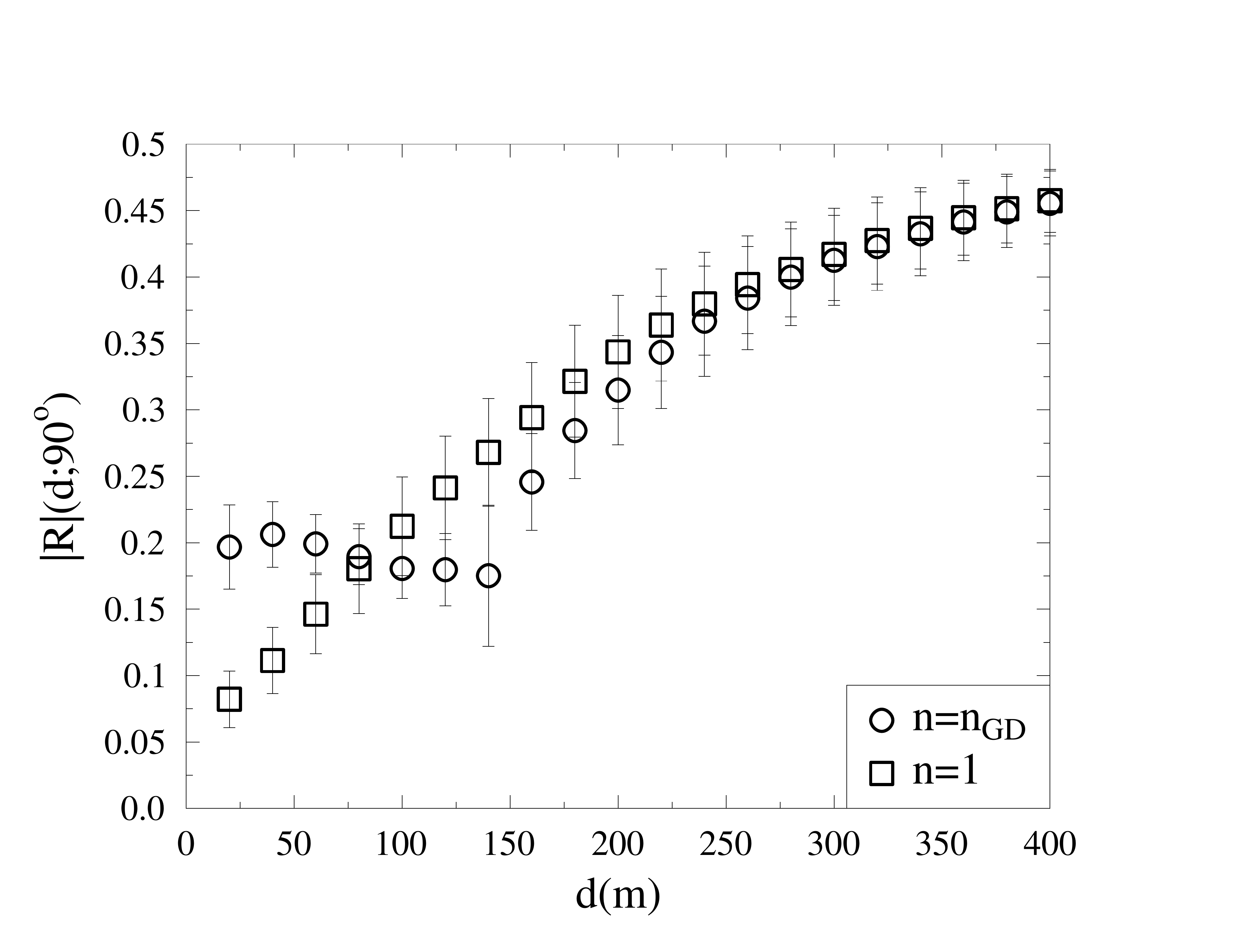}}
\caption{The average value of $|R|$ as defined in~\eqref{R} at an observer angle $\psi=90$ degrees as a function of distance from the shower axis $d$ for a set of 24, $10^{17}$~eV, showers with a proton as primary shower-inducing particle. The shower-inducing particle moves perpendicular to Earth's surface. The squares give $R$ for an index of refraction equal to unity, the circles give $R$ for a realistic index of refraction following the law of Gladstone and Dale given in~\eqref{gd}.}
\figlab{Rvsd}
\end{figure}
\begin{figure}[ht]
\centerline{
\includegraphics[width=.5\textwidth, keepaspectratio]{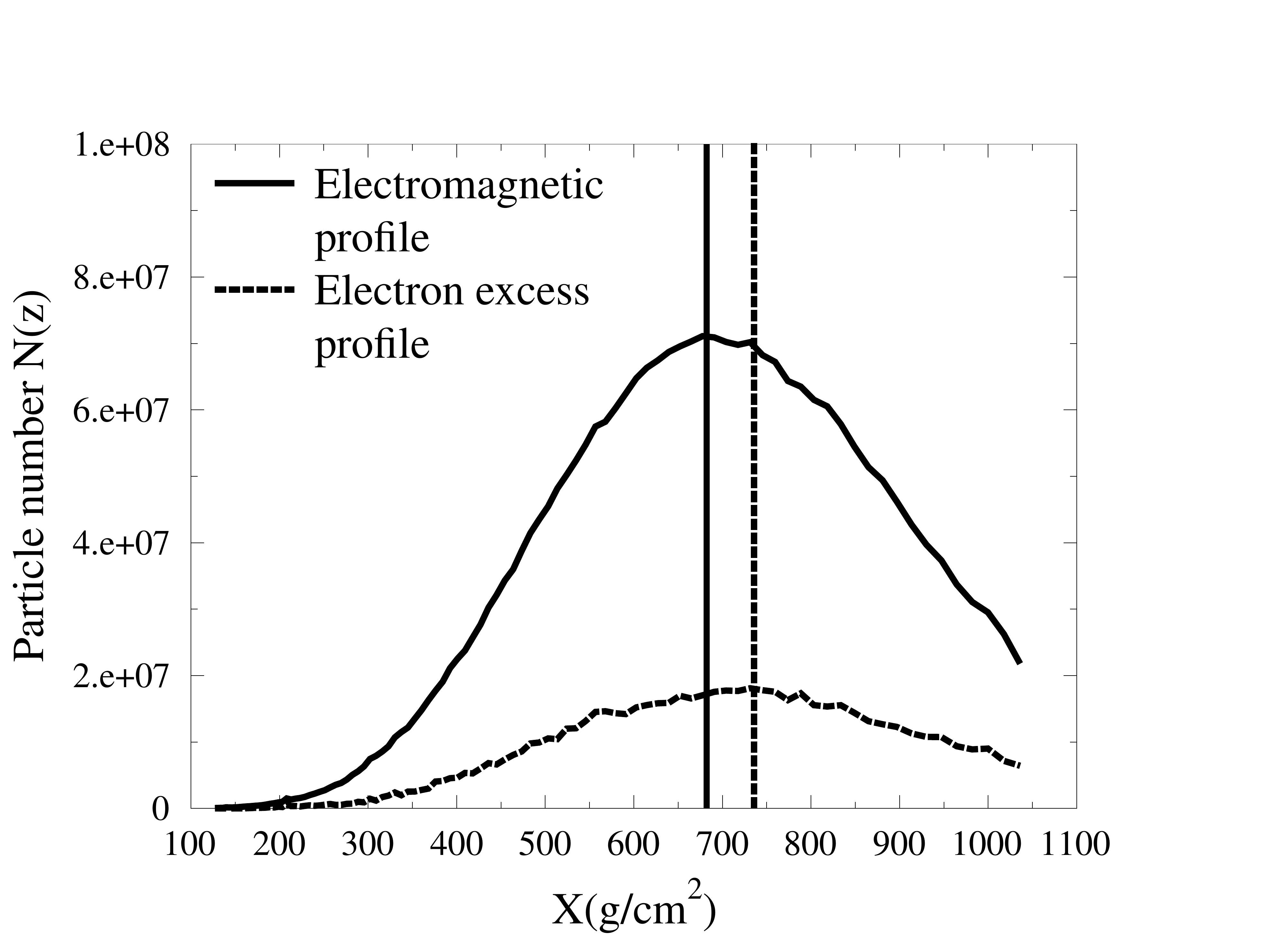}}
\caption{The shower profile given by the full electromagnetic component (full line), and the charge-excess profile (dashed line) as a function of depth along the shower trajectory $X(\mathrm{g/cm^2})$. The vertical lines denote the position of the profile maximum. The charge-excess profile peaks closer to Earth's surface than the shower profile given by the full electromagnetic component.}
\figlab{profile}
\end{figure}

\section{The Lateral Distribution Function:\\The Cherenkov ring}
Due to the extremely thin verticle particle distribution close to the shower axis the electric field contains very high frequency components when the shower maximum is seen at the Cherenkov angle~\cite{Wer12}. Away from the Cherenkov angle the emitted radiation is not seen at a single instant any more and the projection of the shower profile is the determining length scale. This is considered as the `normal' varying current radiation scaling with the derivative of the shower profile and is limited to lower frequencies.
\begin{figure}[!ht]
\centerline{
\includegraphics[width=.5\textwidth, keepaspectratio]{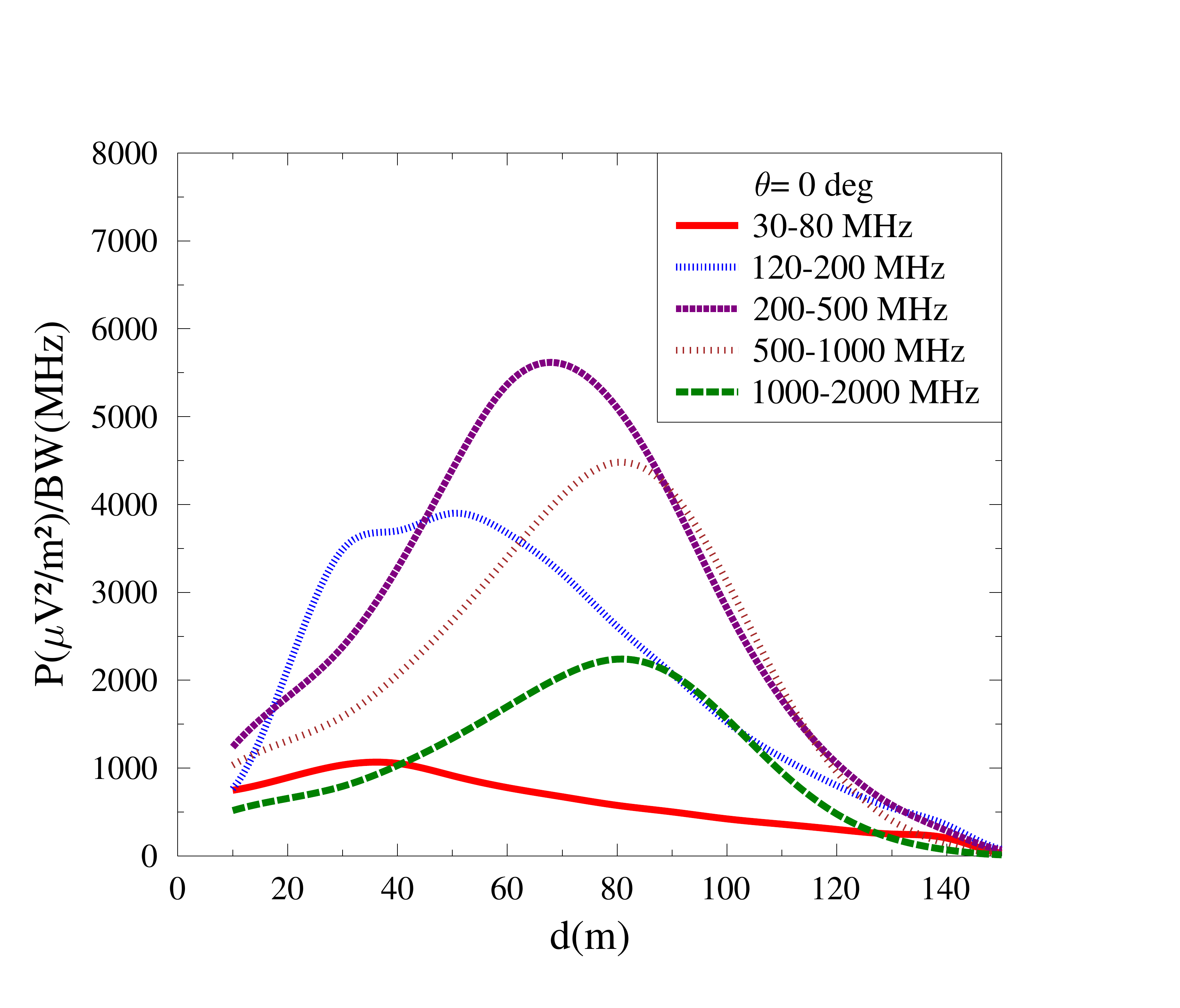}}
\caption{The Lateral Distribution Function (LDF) for different frequency bandwidths. The LDF is given by the intensity of the radio signal divided by the bandwidth as a function of distance to the shower axis. The LDF is calculated for a typical $10^{17}$~eV shower induced by a proton primary particle moving perpendicular to Earth's surface.}
\figlab{ldf_bw}
\end{figure}
This trend for the different frequency components is also seen in~\figref{ldf_bw}. Here the Lateral Distribution Function (LDF) is shown. The LDF is given by the intensity of the radio signal divided by the bandwidth as a function of distance to the shower axis. In~\figref{ldf_bw}, the LDF is shown for different frequency bandwidths. The simulation is done for a typical $10^{17}$~eV proton shower. The LDF at low frequencies is naturally sensitive to emission on a rater long time scale, which is generally given by the `normal' varying current emission and peaks close to the shower axis. For the higher frequencies the LDF becomes more sensitive for Cherenkov emission from close to the shower maximum and peaks further outward. In the $120-200$~MHz band, Cherenkov emission and normal emission compete with each other giving rise to a two peak structure.

\section{Determining $X_{max}$ from the radio signal}
For an index of refraction equal to unity it is shown that one can distinguish between different shower-inducing primary particles~\cite{dVries10,Hue08}. This is due to the fact that at observer distances close to the shower axis the electric field is sensitive to the particle distributions in the shower front which do not differ significantly for proton and iron induced showers. At a large observer distance, on the other hand, the electric field is mainly sensitive to the shower profile, and thus also to $X_{max}(\mathrm{g/cm^2})$, defined by the position where the full electromagnetic component of the shower reaches a maximum. By taking ratios at different observer distances a handle on $X_{max}$, which is closely related to the mass of the initial cosmic ray, is obtained. 

In this section we will show a similar ratio as well as a new, more direct procedure to yield accurate information about $X_{max}$ when Cherenkov effects are taken into account.

In~\cite{dVries12}, it was noticed that the observer position $d=\sqrt{(x^{\perp 1})^2+(x^{\perp 2})^2}$ that corresponds to the Cherenkov angle can be linked to the emission height along the shower axis by,
\beq
d_c=\sqrt{n^2\beta^2-1}\;x^{||}_c\;,
\eqlab{dvsx}
\eeq   
for a constant index of refraction $n$. A more realistic model for the index of refraction is given by the law of Gladstone and Dale given in~\eqref{gd}, where the index of refraction depends on the air density and thus atmospheric height. To test if the relation in~\eqref{dvsx} can be used when the law of Gladstone and Dale is applied, we simulated a set of 100 showers with a $10^{17}$~eV proton as primary shower-inducing particle, and a set of 20 showers with iron as primary shower-inducing particle. The primary shower-inducing particle moves perpendicular to Earth's surface. For each of these simulations we determine the peak of the LDF, $d_p$, and the position of the shower maximum, $x^{||}_m$, which can be easily linked to $X_{max}$. In~\figref{x_max}, we plot $X_{max}$ as a function of $d_p$ for different frequency bandwidths. Showers for which the radio signal peaks below 30 meters have been omitted from~\figref{x_max} and the following analysis.
\begin{figure}[!ht]
\centerline{
\includegraphics[width=.5\textwidth, keepaspectratio]{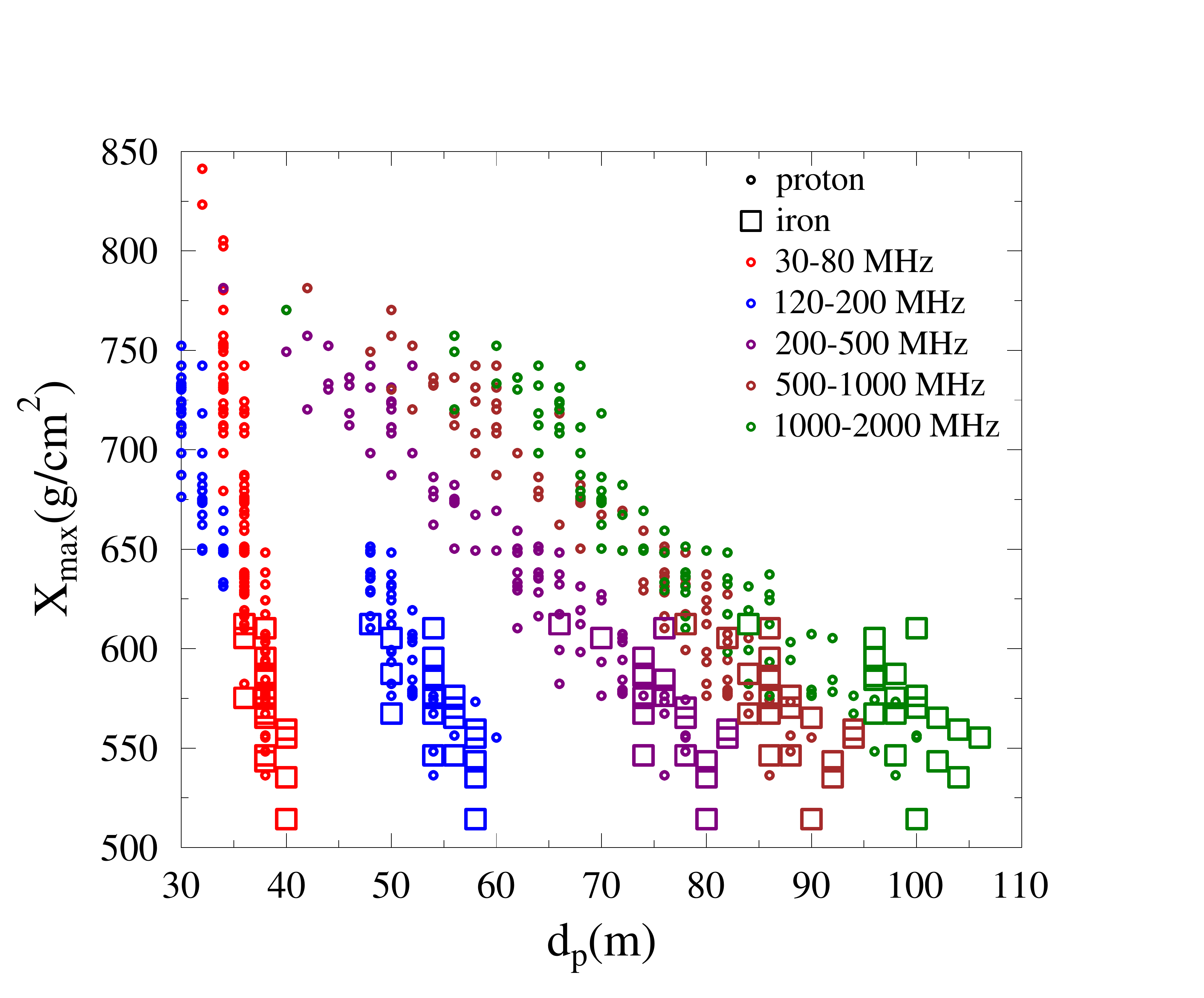}}
\caption{The value of $X_{max}(\mathrm{g/cm^2})$ as a function of $d_p(\mathrm{m})$ the position of the peak in the LDF for different frequency bands for a set of 100 proton (circles) and 20 iron (squares) induced showers.}
\figlab{x_max}
\end{figure}
From~\figref{x_max} it follows that $X_{max}$ can be parameterized as
\beq
X_{max}=a+b\cdot d_p.
\eqlab{Xmax}
\eeq
A fit with~\eqref{Xmax} is sufficient to obtain $X_{max}$ with an accuracy of $10-15\;\mathrm{g/cm^2}$ for the higher frequency bands ($>200$~MHz). It has been tested that this is independent of the energy of the incoming primary particle. In~\figref{x_max_hf} this is shown in the 200-500 MHz band for three different energies of $10^{17}$, $10^{18}$, and $10^{19}$~eV. For each energy a set of 20 showers with a proton as primary shower-inducing particle and a set of 20 showers with an iron atom as primary shower-inducing particle is simulated. The fit parameters $a$, and $b$, and standard deviations are given in Table 1. 
\begin{figure}[!ht]
\centerline{
\includegraphics[width=.5\textwidth, keepaspectratio]{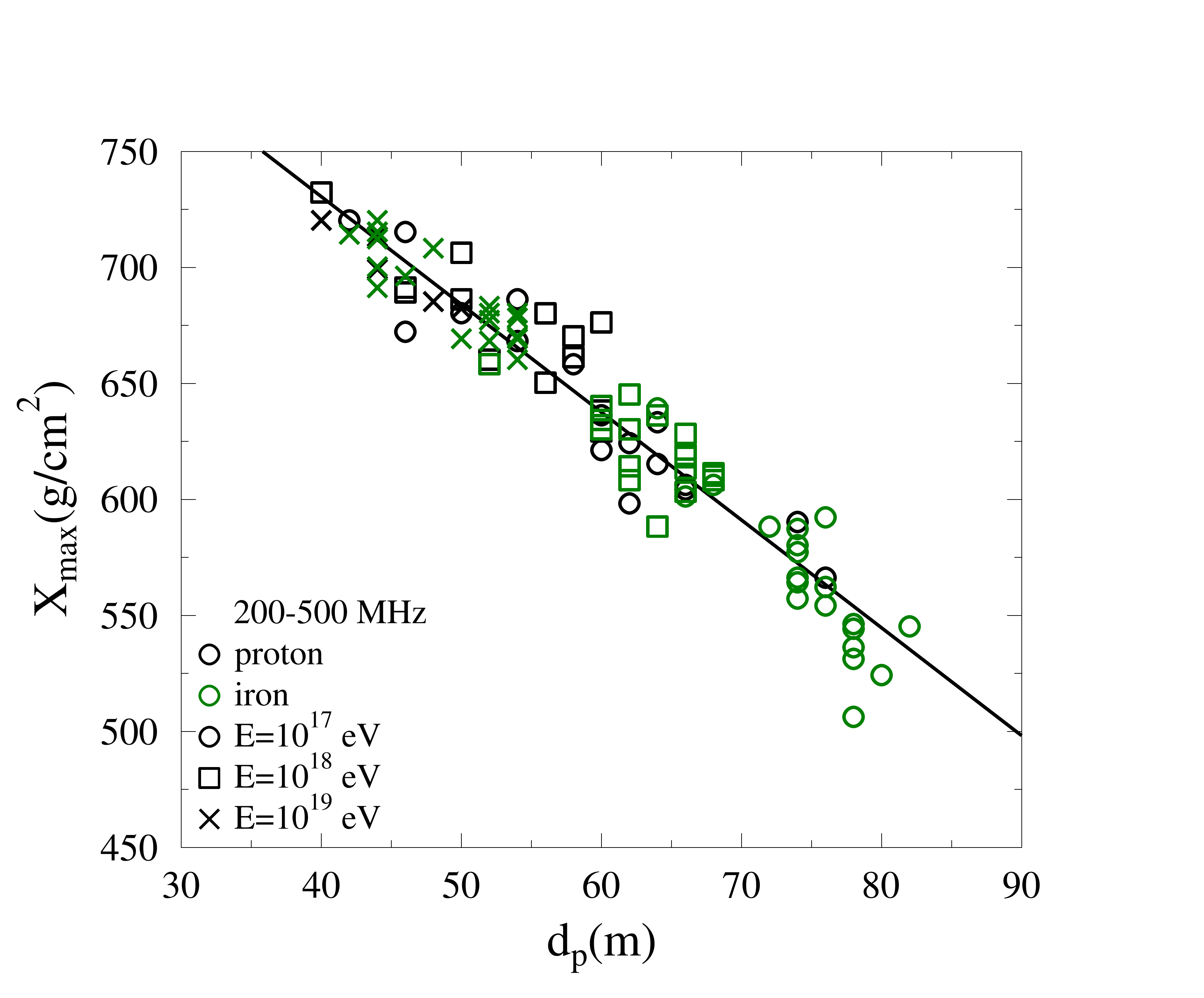}}
\caption{The value of $X_{max}(\mathrm{g/cm^2})$ as a function $d_p(m)$ fitted by~\eqref{Xmax} in the 200-500 MHz band. For three different energies, $10^{17}$~eV, $10^{18}$~eV, and $10^{19}$~eV, a set of 20 proton and 20 iron induced showers has been simulated. The standard deviation of the fit is $14.2\;\mathrm{ g/cm^2}$.}
\figlab{x_max_hf}
\end{figure}
\begin{table*}
\caption{The fit parameters to Eq. 8 for the high frequency bands ($>$ 200 MHz), and Eq. 10 for the low frequency bands ($<$ 200 MHz). The standard deviation of the fits is also given.}
\label{table:1}
\centering
\begin{tabular}{|c | c | c| c|}
\hline
Band-Width & scale parameter & slope parameter & standard deviation ($\mathrm{g/cm^2}$) \\
\hline
30-80 MHz & $\alpha$=161.1 & $\eta$=88.2 & $\sigma=14.4$ \\
\hline
120-200 MHz & $\alpha$=-208.1 & $\eta$=219.9 & $\sigma=10.2$ \\
\hline
200-500 MHz & a=916.4 & b=4.6 & $\sigma=14.2$ \\
\hline
500-1000 MHz & a=898.2 & b=3.7 & $\sigma=11.8$ \\
\hline
1000-2000 MHz & a=860.3 & b=2.9 & $\sigma=12.0$ \\
\hline
\end{tabular}
\end{table*}
From~\figref{x_max} it is clear that for the low frequency bands ($< 200$~MHz), the sensitivity for $X_{max}$ as a function of distance to the shower axis is small. For the $120-200$~MHz band a jump is observed in $X_{max}$ as a function of distance to the shower axis. This jump is related to the two peak behavior in~\figref{ldf_bw}, and is due to the interplay between the normal radiation and Cherenkov emission. Following~\cite{dVries10,Hue08} we plot in~\figref{x_max_lf}, $X_{max}$ as a function of $Q^{30-80}_{300/100}$ which is defined as
\beq
Q^{BW}_{d2/d1}=\frac{P^{BW}(\mu V^2/m^2/MHz;d=d2\;m)}{P^{BW}(\mu V^2/m^2/MHz;d=d1\;m)},
\eqlab{Q}
\eeq
the power in a fixed Band-With (BW) observed at distance $d=d2$~m from the shower axis divided by the power at a distance $d=d1$~m from the shower axis. This analysis is limited to bands at low frequencies ($<$ 200 MHz). One of the simulated showers peakes to close to Earth's surface and has been excluded from the analysis.
\begin{figure}[!ht]
\centerline{
\includegraphics[width=.5\textwidth, keepaspectratio]{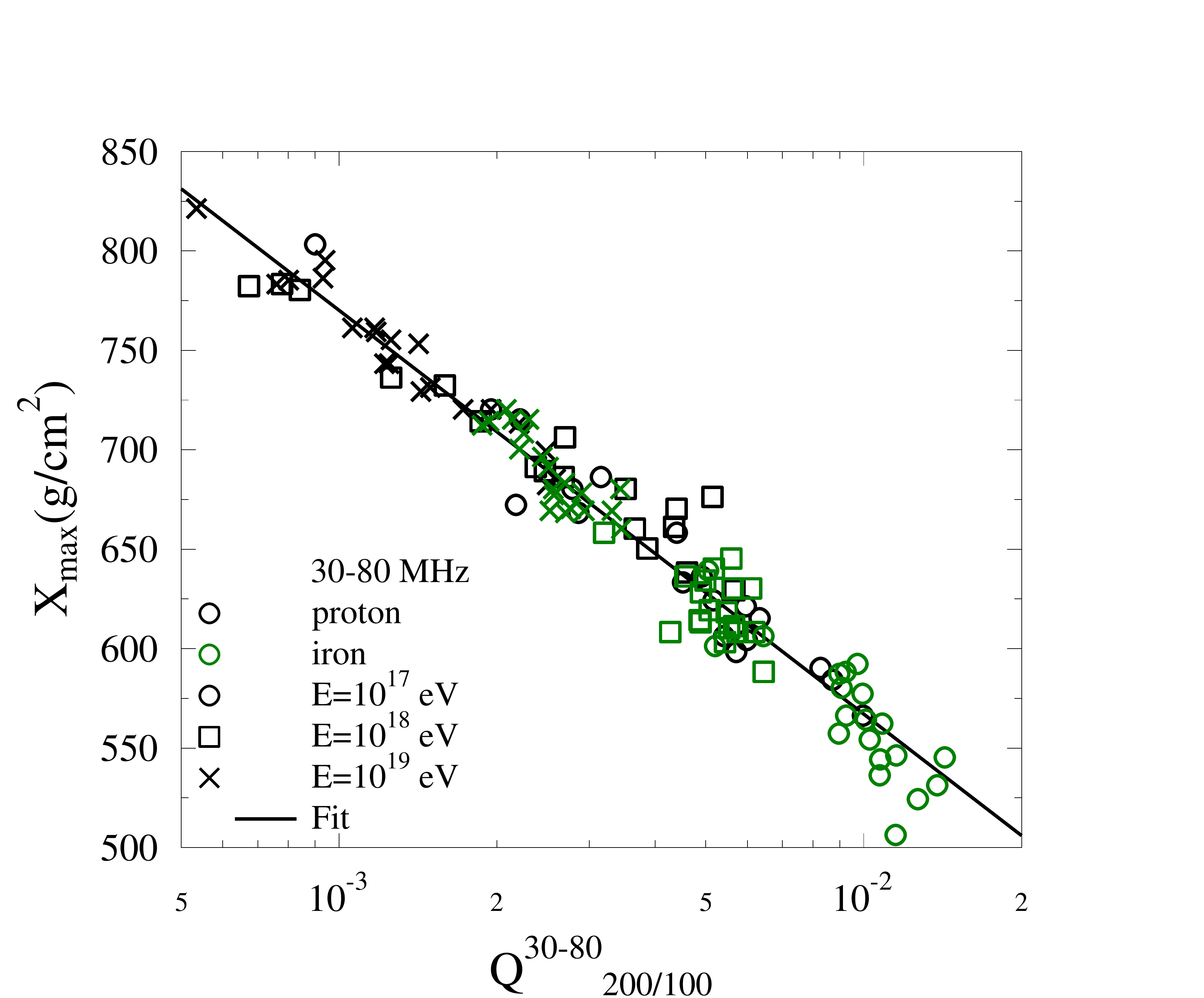}}
\caption{The value of $X_{max}(\mathrm{g/cm^2})$ as a function $Q^{30-80}_{300/100}$ defined by~\eqref{Q} for the 30-80 MHz band. For three different energies, $10^{17}$~eV, $10^{18}$~eV, and $10^{19}$~eV, a set of 20 proton and 20 iron induced showers has been simulated. The standard deviation of the fit is $14.4\;\mathrm{ g/cm^2}$.}
\figlab{x_max_lf}
\end{figure}
The simulations can be fitted with a logarithmic function given as
\beq
X_{max}=\alpha-\eta\log(Q^{BW}_{d2/d1}),
\eeq
obtaining an accuracy of $\sim10-15\;\mathrm{g/cm^2}$. The fit parameters $\alpha$, and $\eta$, and the standard deviations are given in Table 1 for the different low frequency bands ($<$ 200 MHz).\\

It should be noted that the obtained results are for a simplified geometry of showers moving perpendicular to Earth's surface. Also instrumental errors have not been included into the analysis. In comparison, the Fluorescence detection technique employed at the Pierre Auger Observatory~\cite{Aug11}, determines the shower maximum with an approximate accuracy of $20\;\mathrm{g/cm^2}$. 

\section{Conclusions}
In this article we have shown that radio emission from cosmic-ray-induced air showers can be used to obtain relevant information about the emission mechanisms and the composition of the cosmic-ray spectrum. The influence of Cherenkov effects on the emission mechanisms, the geomagnetic emission due to the deflection of charged leptons in Earth's magnetic field and the charge-excess emission due to a net negative electron excess in the shower front, is investigated. This is studied by means of the $R$ parameter which is a measure for the charge-excess component in the emission. It is shown that since the polarization is not affected by Cherenkov effects, the qualitative behavior of $R$ as a function of observer angle does not change. Nevertheless, the behavior of $R$ as a function of observer distance to the shower axis is shown to be sensitive to Cherenkov effects. It is shown that these effects can be used as a direct measure of the relative position of the maximum number of particles, $X_{max}$, for the full shower profile given by the total electromagnetic component with respect to the charge-excess profile given by the electron excess in the shower front.

Cherenkov effects determine the structure of the lateral distribution function (LDF) defined by the signal strength as a function of distance to the shower axis. For the high frequency bands ($>200$~MHz), it is shown that the peak position in the LDF can be linked to $X_{max}$ with an  accuracy of $10-15\;\mathrm{g/cm^2}$. For the low frequency bands ($<200$~MHz), we use the power ratio, $Q^{BW}_{d2/d1}$, given by the power at distance $d2$ divided by the power at distance $d1$ as a measure for $X_{max}$. With this method a similar accuracy of $10-15\;\mathrm{g/cm^2}$ is obtained. These values are obtained for a simplified geometry of vertical air showers, and instrumental errors have not been included into the analysis.

\section{Acknowledgment}
This work is part of the research program of the 'Stichting voor Fundamenteel
Onderzoek der Materie (FOM)', which is financially supported by the 'Nederlandse
Organisatie voor Wetenschappelijk Onderzoek (NWO)'.

\appendix

\end{document}